\documentclass[preprint,pra,aps]{revtex4}
\usepackage{amssymb}

\usepackage{graphicx}

\begin{document}

\title{Correction to the Casimir force due to the anomalous skin effect}
\author{R. Esquivel}
\email[]{raul@fisica.unam.mx}
\affiliation{Instituto de Fisica,
Universidad Nacional Aut\'{o}noma de M\'{e}xico, Apartado Postal
20-364, DF 01000 M\'{e}xico, Mexico}
\author{V. B. Svetovoy}
\email[]{V.B.Svetovoy@el.utwente.nl}
\thanks{On leave from Yaroslavl University, Yaroslavl, Russia}
\affiliation{Transducers Science and Technology Group, EWI,
University of Twente, P.O. 217, 7500 AE Enschede, The Netherlands}
\date{\today}

\begin{abstract}
The surface impedance approach is discussed in connection with the
precise calculation of the Casimir force between metallic plates.
It allows to take into account the nonlocal connection between the
current density and electric field inside of metals. In general, a
material has to be described by two impedances $Z_{s}(\omega,q)$
and $Z_{p}(\omega,q)$ corresponding to two different polarization
states. In contrast with the approximate Leontovich impedance they
depend not only on frequency $\omega$ but also on the wave vector
along the plate $q$. In this paper only the nonlocal effects
happening at frequencies $\omega<\omega_{p}$ (plasma frequency)
are analyzed. We refer to all of them as the anomalous skin
effect. The impedances are calculated for the propagating and
evanescent fields in the Boltzmann approximation. It is found that
$Z_p$ significantly deviates from the local impedance as a result
of the Thomas-Fermi screening. The nonlocal correction to the
Casimir force is calculated at zero temperature. This correction
is small but observable at small separations between bodies. The
same theory can be used to find more significant nonlocal
contribution at $\omega\sim\omega_p$ due to the plasmon
excitation.
\end{abstract}


\maketitle

\section{Introduction\label{Sec1}}

The Casimir force \cite{Cas} between uncharged metallic plates,
predicted in 1948 as a quantum electrodynamics effect, only
recently became a subject of systematic experimental
investigation. The reason is that nowadays with the development of
microtechnologies a reliable control of the separations between
bodies smaller than 1 $\mu m$ became possible. A variety of
methods have been used to measure the force. In the torsion
pendulum experiment \cite{Lam1}, first in the modern series, the
force between a sphere and a plate covered with gold was measured
with the accuracy of 5\%. A significant progress was achieved in
the atomic force microscope (AFM) experiments \cite{Moh1,Moh2},
where the sphere was attached to a cantilever. The force was
deduced from the cantilever bending when the plate was approaching
the sphere. In this experiment the force was measured with a 1\%
precision. The best result using AFM \cite{Moh2} was found when
the sphere and the plate were covered with gold and special care
was taken to control the surface roughness of metal. The same
precision was reached in the crossed cylinders experiment
\cite{Ederth}, where extremely smooth gold films were used.
Sophisticated microelectromechanical system (MEMS) \cite{Chan} was
used to measure the force between the gold plated sphere and a
suspended paddle. It demonstrated the nonlinear behavior of the
mechanical oscillator due to the Casimir force. The only
experiment, where the force was measured in the plate-plate
geometry \cite{Bressi}, was done using an oscillating beam whose
resonance frequency changed in response to the force. Up to date,
the most precise experiment \cite{Decca1,Decca2} explored the MEMS
device similar to that in Ref. \cite{Chan}. The precision was
improved due to the use of the dynamical method. Additionally the
change in the resonance frequency of the mechanical oscillator was
measured using the phase jump instead of resonance behavior of the
amplitude. In this way the force  was found with the relative
accuracy of 0.25\% \cite{Decca1,Decca2,Iannuzzi}.

To draw any conclusion from the experiments one has to predict the
force theoretically with the precision comparable with the
experimental errors. In its original form, the Casimir force
\cite{Cas}

\begin{equation}
F_{c}\left( a\right) =-\frac{\pi ^{2}}{240}\frac{\hbar c}{a^{4}}
\label{Fc}
\end{equation}

\noindent was calculated between the ideal metals. It depends only
on the fundamental constants and the distance between the plates
$a$. The force between real materials, described by its dielectric functions $\varepsilon \left( \omega \right) $,
was deduced for the first time by Lifshitz \cite{Lif}.  Corrections to Eq. (\ref{Fc}) can be  quite
significant at small separation between bodies. To calculate the
force with high precision the Lifshitz formula is used with the
optical data taken from handbooks \cite{HB1,HB2}. The data are
available only up to some low cut-off frequency  $\omega _{cut}$.
For good metals such as $Au,$ $Al,$ $Cu$ the data can be
extrapolated to lower frequencies with the Drude dielectric
function

\begin{equation}
\varepsilon \left( \omega \right) =1-\frac{\omega _{p}^{2}}{\omega
\left( \omega +i\omega _{\tau }\right) },  \label{Drude}
\end{equation}

\noindent which includes two parameters: the plasma frequency
$\omega _{p}$ and the relaxation frequency $\omega _{\tau }$.
These parameters can be extracted from the optical data at the
lowest accessible frequencies. In this way the force has been
calculated \cite{Lam2,BS0,LR,SL2,KMM} with the highest possible
precision. There is some disagreement between the results of
different authors connected with the choice of the relaxation
frequency $\omega _{\tau }$ \cite{SL2}. This frequency can be
found by fitting the low frequency optical data with Eq.
(\ref{Drude}) \cite{LR} or extracted from the bulk material
resistivity \cite{KMM}. These details are important at small
separations.

In Refs. \cite{SL1,SL2} it was stressed that using the handbook
optical data one finds not the actual force but rather the upper
limit on the Casimir force. The reason is that the handbooks
comprise the data for the best samples; any material imperfection
will reduce the reflectivity and, as a result, the force will be
smaller. In the experiments the force is measured between the
bodies covered with a metal. The metal is deposited on a substrate
with the evaporation or sputtering technics \cite{Moh2,Decca1}.
The resulting film thickness is typically in the range
$100-200\,nm$. It was already noted \cite{SL1,SL2} that the
optical properties of the films can deviate significantly from
those of the bulk material. The main reasons for these deviations
such as voids in the films and electron scattering on the grain
boundaries were indicated recently \cite{S} and the methods to
estimate the effects were outlined. For the AFM \cite{Moh2} and
MEMS \cite {Decca2} experiments the influence of these effects on
the force was estimated on the level of 2\%. Additional work from
the theoretical and experimental sides has to be done to refine
these estimates.

An alternative way to calculate the Casimir force using the
surface impedance of metals instead of the dielectric function was
discussed \cite {BKR,SL4,GKM} in the literature. The general
formula for the Casimir force in this approach was deduced for the
first time in Ref. \cite{MT}. It is the same Lifshitz formula but
the reflection coefficients are expressed in terms of the surface
impedance. In the cited papers the approximate Leontovich
impedance \cite{LL8} depending only on frequency was used. The
hope was that in this approach it will be possible to resolve the
long standing problem with the temperature correction to the force
\cite {BS1,SL2,BGKM,GLR,Lam3,SL3,KM,BAH,HBAM}. However, although
the impedance approach itself seems reasonable and well motivated,
the use of the approximate impedance for the Casimir force
happened to be unjustified. The Leontovich impedance is well
suited for the propagating photons impinging on a metal but for
the Casimir problem the exact impedance has to be used since the
important contribution in the force comes from the evanescent
electromagnetic field. The exact impedance was shown to give the
same result for the force as that in the dielectric function
approach \cite{MVE}.

The surface impedance is the only way to describe the interaction
between the electromagnetic field and metal in the case when the
relation between current and electric field in the metal becomes
nonlocal. For example, at low temperatures the mean free path of the
electrons in metals becomes larger than the field penetration
depth, and the relation between the current and
field becomes nonlocal at low frequencies \cite{LP10,Abr} and the
anomalous skin effect is realized. In connection with the Casimir
force the contribution of this effect in the temperature
correction was discussed in Ref.  \cite {SL4}. On the other hand at high
frequencies $\omega \sim \omega _{p}$ the charge density
fluctuations can propagate in the material (plasmons) taking away
the energy from the incoming field. This is also an example of
nonlocal effect which was shown to give a significant correction
to the Casimir force \cite{EVM} at small separations between
bodies. In the nonlocal case both time and space dispersion happen
and the dielectric function depends not only on frequency $\omega
$ but also on the wave vector $k$. Actually in this case one has
to separate two dielectric functions: the longitudinal function
$\varepsilon _{l}\left( \omega ,k\right) $, which describes
material response to the longitudinal (in respect to $k$) electric
field, and the transverse dielectric function $\varepsilon
_{t}\left( \omega ,k\right) $ describing the response to the
transverse field. A systematic way to calculate the surface
impedances via the dielectric functions and explicit expressions
for these functions were given in a series of classical papers by
Kliewer and Fuchs \cite{KF1,KF2,KF3}.

The aim of this paper is to provide the basis
for systematic investigation of the corrections to the Casimir
force due to the nonlocal effects. Here we consider only the
minor corrections which appear at low frequencies $\omega <<\omega
_{p}$ but introduce a general approach which is true at any
frequencies. This approach is not new in the condensed matter
physics but it has never  been discussed before in connection with
the Casimir effect.  

We deliberately do not consider here the correction to the force
in the nonzero temperature case though our conclusions about the
low frequency behavior of the impedances for two polarization
states will be important for the discussion of the temperature
correction.

The paper is organized as follows. In Sec. \ref{sec2} the main
definitions are introduced allowing to calculate the impedances
for two polarization states via the longitudinal and transverse
dielectric functions. The explicit expressions for the nonlocal
dielectric functions in the Boltzmann approximation are given. The
impedances are calculated first for propagating fields to compare
the calculations with the known results. Then we discuss the
continuation procedure to the range of evanescent fields and
calculate the impedances at imaginary frequencies to put them
later in the Lifshitz formula. The low frequency behavior of these
impedances important for the thermal correction to the Casimir
force is discussed specifically. In Sec. \ref{sec3} the actual
calculations of the correction to the Casimir force due to
nonlocal effects are presented for plate-plate and sphere-plate
geometry. The discussion and concluding remarks are given in Secs.
\ref{sec4} and \ref{sec5}.

\section{Surface impedances of metals\label{sec2}}

The anomalous skin effect was incorporated into the general theory
of the optical properties of metals with the detailed paper by
Reuter and Sondheimer \cite{RS}; the qualitative description was
given earlier by Pippard \cite{Pip}. Pippard was the first to
point out that, in general, the electric field inside a metal
cannot be considered as spatially constant. In general,  the
connection between the current and the field becomes nonlocal. The
current is given by a definite integral involving the values of
the electric field at all points in the metal, and the Maxwell's
equations therefore lead to integro-differential equation from
which the electric field has to be determined. The expression for
the current was deduced \cite{RS} at the conditions that the
electrons can be considered as quasi-free and the collision
mechanism can always be described in terms of a mean free path $l$
or, alternatively, a relaxation frequency $\omega _{\tau }$. The
mean free path was assumed to be independent on the direction of
motion. An additional assumption about the electron reflection off
the surface was introduced by Pippard: a fraction $P$ of the
electrons arriving at the surface is scattered specularly, while
the rest are scattered diffusely. Only normal incidence of the
electromagnetic field on the metal was considered in Ref.
\cite{RS}, however, it was stressed in the Kliewer and Fuchs paper
\cite{KF1} that the theory of anomalous skin effect can be
considered complete only when arbitrary incidence is taken into
account.

The relation between the current density and the electric field
can be found solving the Boltzmann equation if the free electrons
and field wavelengths large compared to the wavelength of an
electron at the Fermi surface. For the description of the
anomalous skin effect it is a valid approximation. In general,
this relation can be found in different approximations using the
linear response theory \cite{Wang}.

\subsection{Propagating waves}

\begin{figure}
\includegraphics[width=8.6cm]{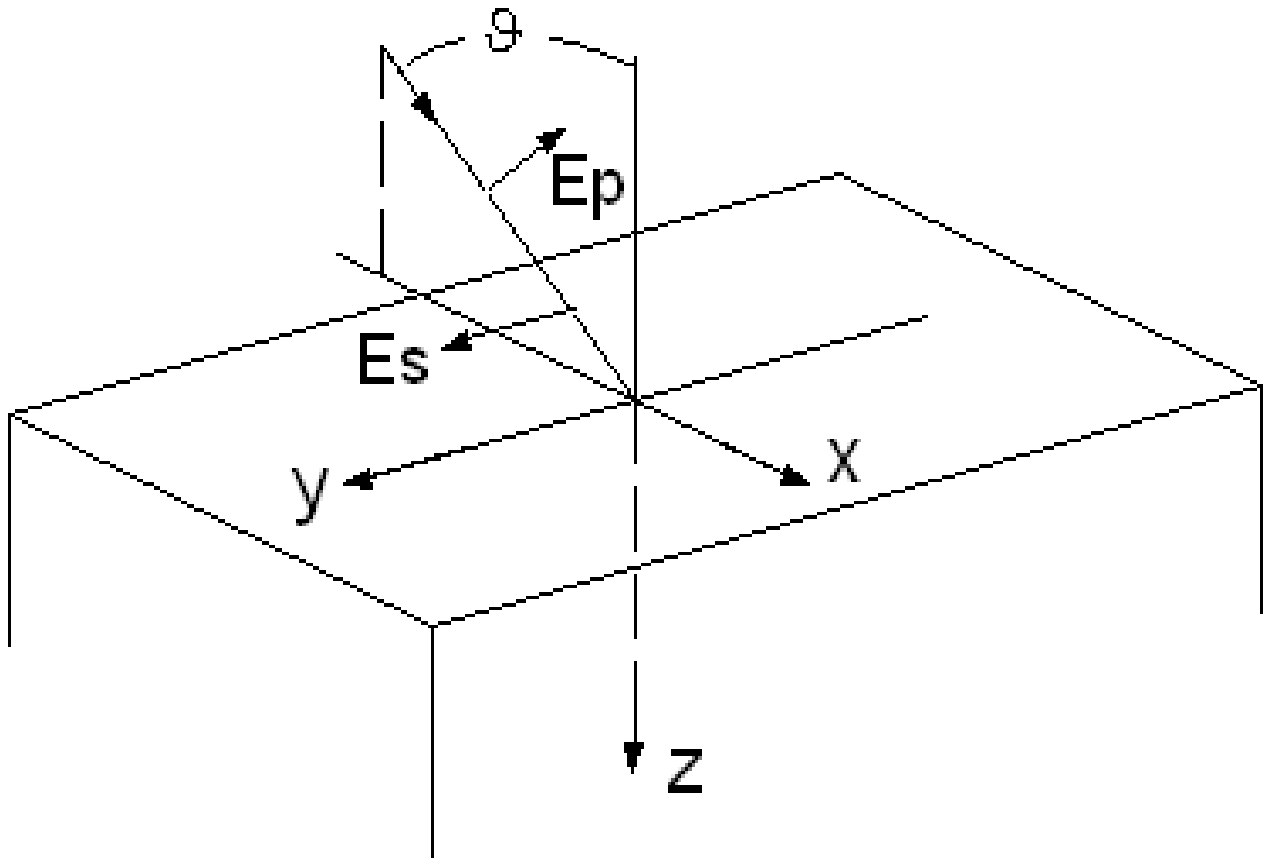}
\caption{Choice of the coordinate system for the incoming wave.
The angle of incidence is given by $\vartheta$. The electric
fields in the $p$ polarized wave $E_p$ and $s$ polarized wave
$E_s$ are shown.} \label{fig1}
\end{figure}

Following Kliever and Fuchs \cite{KF1} we consider a plane wave of
angular frequency $\omega $  incident from vacuum at an angle
$\vartheta $ upon the surface of metallic half-space. The geometry,
together with the choice of the coordinate system, is shown in Fig.
\ref{fig1}. One can separate two polarization states for the wave.
For the $s$ polarized wave the electric field is directed in the $
y $ axis, while for the $p$ polarized wave the electric field is
in the $x-z$ plane and has nonzero $z$ component. For clarity let
us sketch out how the specific expressions for the surface
impedances have been deduced in Ref. \cite{KF1}.

For the $s$ polarized wave the field can be written in the form
${\bf E}=E_y(z)\exp (-i\omega t+ik_{x}x) \textbf{n}_y$, where
$\textbf{n}_y$ is the unit vector along the $y$ axis and
$k_{x}=\left( \omega /c\right) \sin \vartheta $ is the
$x$-component of the wave vector in the incoming wave. Since this
component will play significant role in what follows, we will use
for it also a special notation $q\equiv k_{x}$, which is settled
in the field of the Casimir force. The $z$ dependence of the electric
field ${\bf E}=(0,E_{y},0) $ for the $s$ wave inside of metal can
be described with the Maxwell equation

\begin{equation}
\frac{d^{2}E_{y}}{dz^{2}}-k_{x}^{2}E_{y}+\frac{\omega
^{2}}{c^{2}}D_{y}=0, \label{Maxeq}
\end{equation}

\noindent where ${\bf D}$ is the displacement field. This equation
is valid for $z>0$. Taking the Fourier transform of Eq.
(\ref{Maxeq}) over the $z$ coordinate, we obtain

\begin{equation}
-(k_{x}^{2}+k_{z}^{2}){\cal E}_{y}+\frac{\omega ^{2}}{c^{2}}{\cal D}_{y}=
\frac{dE_{y}\left( +0\right) }{dz}-\frac{dE_{y}\left( -0\right) }{dz},  \label{MaxFour}
\end{equation}

\noindent where ${\cal E}$ and ${\cal D}$ are the Fourier
transformed fields defined as

\begin{equation}
{\cal E}(k_z)=\int\limits_{-\infty }^{\infty }dzE\left( z\right)
e^{-ik_{z}z} \label{Four}
\end{equation}

\noindent and similarly for ${\cal D}$. Equation (\ref{MaxFour})
describes the behavior of the electric field in an infinitely
extended medium. Furthermore,  the right hand side in Eq.
(\ref{MaxFour}) is undefined until we find a relation between the
two derivatives at the surface.  This relation involves describing
or modelling the surface. In this work we assume that the
electrons scatter elastically from the surface (specular
reflection).  This assumption is equivalent to assuming an
infinitely extended medium, needed to obtain Eq. (\ref{MaxFour}),
since an electron bouncing from the surface cannot be
distinguished from an electron coming from a fictitious medium on
the vacuum side.   This is  taken into account imposing the
symmetry requirements

\begin{equation}
E_{y}\left( z\right) =E_{y}\left( -z\right) ,\quad D_{y}\left(
z\right) =D_{y}\left( -z\right) .  \label{sym}
\end{equation}

\noindent Thus, using  the Maxwell equation
\begin{equation}
\frac{dE_{y}(+0)}{dz}-\frac{dE_{y}(-0)}{dz}=-i\left( \omega /c\right) H_{x},
\end{equation}

\noindent and from Eq. (\ref{MaxFour}) one finds

\begin{equation}
\frac{{\cal E}_{y}}{H_{x}\left( +0\right) }=-\frac{2i\omega }{c}\frac{1}{%
\left( \omega ^{2}/c^{2}\right) \varepsilon _{t}-k^{2}},\quad
k^{2}=k_{x}^{2}+k_{z}^{2}.  \label{EHrel}
\end{equation}

\noindent The inverse Fourier transform of this equation evaluated
at $z=0$ gives the surface impedance for $s$ polarization:

\begin{equation}
Z_{s}\left( k_{x},\omega \right) \equiv -\frac{E_{y}\left( +0\right) }{%
H_{x}\left( +0\right) }=\frac{i}{\pi }\frac{\omega
}{c}\int\limits_{-\infty }^{\infty }\frac{dk_{z}}{\left( \omega
^{2}/c^{2}\right) \varepsilon _{t}-k^{2}}.  \label{Zs}
\end{equation}

\noindent For $p$ polarization the problem is slightly more
complicated since we have two non-vanishing components of the
electric field $E_x$ and $E_z$. Following a similar line of
reasonings as before,  the impedance for $p$ polarization is
obtained as:

\begin{equation}
Z_{p}\left( k_{x},\omega \right) \equiv \frac{E_{x}\left( +0\right) }{%
H_{y}\left( +0\right) }=\frac{i}{\pi }\frac{\omega
}{c}\int\limits_{-\infty }^{\infty }\frac{dk_{z}}{k^{2}}\left[
\frac{k_{x}^{2}}{\left( \omega ^{2}/c^{2}\right) \varepsilon
_{l}}+\frac{k_{z}^{2}}{\left( \omega ^{2}/c^{2}\right) \varepsilon
_{t}-k^{2}}\right] .  \label{Zp}
\end{equation}

\noindent It is natural that both the longitudinal $\varepsilon
_{l}\left( {\bf k },\omega \right) $ and transverse $\varepsilon
_{t}\left( {\bf k},\omega \right) $ dielectric functions
contribute to $Z_{p}$ because in the $p$ wave the electric field
has both components.

Since the impedance approach caused recently some confusion in the
field of the Casimir force \cite{BKR,SL4,GKM}, a few comments
concerning the impedances Eq. (\ref{Zs}) and Eq. (\ref{Zp}) should
be made. First, there is not one but two impedances corresponding
two different polarizations. Second, the impedances depend not
only on the frequency but also on the wave vector along the metal
surface $q=k_{x}=\left( \omega /c\right) \sin \vartheta $. Only
for the normal incidence the impedances for $p$ and $s$ polarized
waves coincide and depend only on frequency. Third, no specific
assumptions about the dielectric functions $ \varepsilon _{l}$ and
$\varepsilon _{t}$ were made in the derivation of Eq. (\ref {Zs})
and Eq. (\ref{Zp}). In particular, the local functions
$\varepsilon _{l}\left(0,\omega \right) =\varepsilon
_{t}\left(0,\omega \right) =\varepsilon \left( \omega \right) $
can be used. In this case the integrals can be easily calculated
to find so called classical or local impedances

\begin{equation}
Z_{s}^{loc}\left( q,\omega \right) =\frac{1}{\sqrt{\varepsilon
\left( \omega \right) -\left( cq/\omega \right) ^{2}}},\quad
Z_{p}^{loc}\left( q,\omega \right) =\frac{\sqrt{\varepsilon \left(
\omega \right) -\left( cq/\omega \right) ^{2}}}{\varepsilon \left(
\omega \right) }.  \label{loc}
\end{equation}

\noindent These expressions completely coincide with those
introduced in Refs. \cite{MVE,EVM} and, as was shown there,
exactly reproduce the Casimir force in the dielectric function
approach. The Leontovich impedance used in Refs. \cite{BKR,SL4,
GKM} from the beginning was introduced as the approximate one
\cite{LP10,Abr} for applications in radiophysics. For the
propagating waves it really has sense because $cq/\omega \leq 1$
but for metals in the microwave range $\left| \varepsilon \left(
\omega \right) \right| \gg 1$. So one can neglect the wave vector
along the plates ($q=k_{x}$) in the impedances to get just one
frequency dependent function. However, in the Casimir force
significant contribution comes from the evanescent fields for
which $cq/\omega >1$. In this case the Leotovich approximation
fails especially in the limit $\omega \rightarrow 0$, which is
important for the analysis of the temperature correction.

The dielectric functions were found \cite{KF1} solving the
Boltzmann equation and the result is the following

\begin{equation}
\varepsilon _{t}\left( k,\omega \right) =1+\chi _{IB}\left( \omega \right) -%
\frac{\omega _{p}^{2}}{\omega \left( \omega +i\omega _{\tau }\right) }%
f_{t}\left( u\right) ,  \label{etdef}
\end{equation}

\begin{equation}
\varepsilon _{l}\left( k,\omega \right) =1+\chi _{IB}\left( \omega \right) -%
\frac{\omega _{p}^{2}}{\omega \left( \omega +i\omega _{\tau }\right) }%
f_{l}\left( u\right) ,  \label{eldef}
\end{equation}

\noindent where the phenomenological susceptibility $\chi
_{IB}\left( \omega \right) $ was introduced to describe the
interband transitions since these processes are beyond the
quasi-free electron model. The functions $ f_{t,l}\left( u\right)
$ taking into account nonlocal effects are defined in the
following way

\begin{equation}
f_{t}\left( u\right) =\frac{3}{2u^{3}}\left[ u-\frac{1}{2}\left(
1-u^{2}\right) \ln \left( \frac{1+u}{1-u}\right) \right] ,
\label{ftdef}
\end{equation}

\begin{equation}
f_{l}\left( u\right) =\frac{3}{u^{3}}\left[ -u+\frac{1}{2}\ln \left( \frac{%
1+u}{1-u}\right) \right] \cdot \left[ 1+i\frac{\omega _{\tau }}{\omega }-%
\frac{i}{2u}\frac{\omega _{\tau }}{\omega }\ln \left( \frac{1+u}{1-u}\right) %
\right] ^{-1},  \label{fldef}
\end{equation}

\noindent where the variable $u$ responsible for the nonlocal
effects is

\begin{equation}
\qquad u=\frac{v_{F}k}{\omega +i\omega _{\tau }},  \label{udef}
\end{equation}

\noindent and $v_{F}$ is the Fermi velocity. In the local limit
$k\rightarrow 0 $ both functions (\ref{etdef}) and (\ref{eldef})
reduce to the local dielectric function (Drude plus interband
transitions)

\begin{equation}
\varepsilon \left( \omega \right) =1+\chi _{IB}\left( \omega \right) -\frac{%
\omega _{p}^{2}}{\omega \left( \omega +i\omega _{\tau }\right) }.
\label{eps}
\end{equation}

The function $\varepsilon _{t}$ was found first by Reuter and
Sondheimer \cite{RS}. Since these authors considered only the
normal incidence, in their work there was no $\varepsilon _{l}$.
This function appears at non-normal incidence. In the $p$ wave
there is the normal field component $ E_{z}$ giving rise to charge
fluctuations to which the system responds via the longitudinal
dielectric function. It should be mentioned that with the charge
fluctuations the relaxation of the perturbed electron distribution
toward equilibrium was chosen to the local state of charge
imbalance but not to the uniform distribution. The denominator in
Eq. (\ref{fldef}) describes this effect.

The equations (\ref{etdef})-(\ref{udef}) are used in the optics of
metals \cite{Book2} to predict the reflectance or absorptance of
the materials. It is easy to find the reflection amplitudes
$r_{s}$ and $r_{p}$ for $s$ and $p$ polarizations expressed via
the impedances as:

\begin{equation}
r_{s}=\frac{\frac{\omega }{c}-Z_{s}\sqrt{\frac{\omega ^{2}}{c^{2}}-q^{2}}}{%
\frac{\omega }{c}+Z_{s}\sqrt{\frac{\omega ^{2}}{c^{2}}-q^{2}}},\quad r_{p}=%
\frac{\sqrt{\frac{\omega ^{2}}{c^{2}}-q^{2}}-\frac{\omega }{c}Z_{p}}{\sqrt{%
\frac{\omega ^{2}}{c^{2}}-q^{2}}+\frac{\omega }{c}Z_{p}}.
\label{refl}
\end{equation}

\noindent The reflectance and absorptance are given by:

\begin{equation}
R_{s,p}=\left| r_{s,p}\right| ^{2},\quad A_{s,p}=1-\left|
r_{s,p}\right| ^{2}.  \label{RA}
\end{equation}

In what follows we will use dimensionless variables, which are
more convenient for numerical calculations. We define

\begin{equation}
\Omega =\frac{\omega }{\omega _{p}},\quad Q=\frac{cq}{\omega
_{p}},\quad \gamma =\frac{\omega _{\tau }}{\omega _{p}}.
\label{var}
\end{equation}

\begin{figure}
\includegraphics[width=8.6cm]{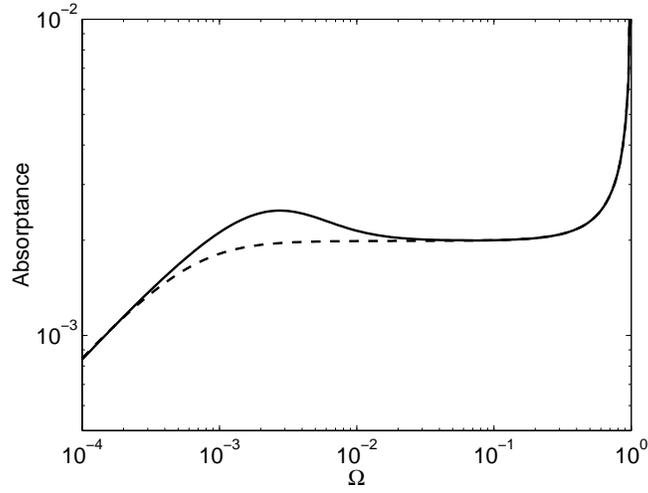}
\caption{Absorptance as a function of the dimensionless frequency
$\Omega$ at normal incidence $\vartheta=0$. Local and nonlocal
cases are represented by the dashed and solid lines, respectively.
At $\vartheta=0$ there is no difference between $s$ and $p$
polarizations. Parameters were chosen as in Ref.
\protect\cite{KF1}: $\protect\gamma=1\cdot10^{-3}$,
$v_F=0.85\cdot10^8\ cm/s$.} \label{fig2}
\end{figure}

\begin{figure}
\includegraphics[width=8.6cm]{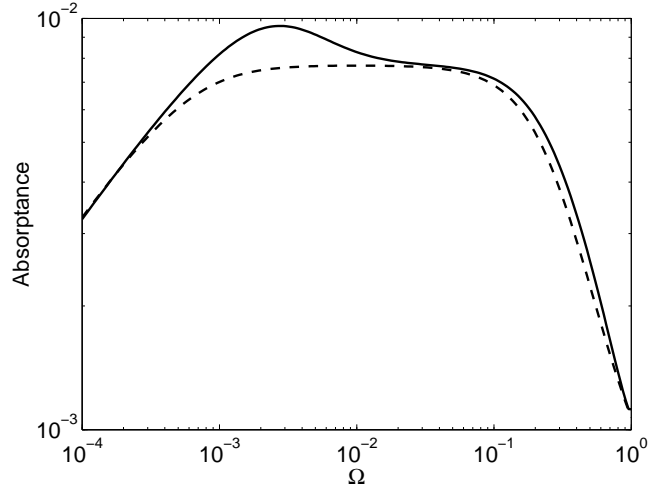}
\caption{Absorptance as a function of $\Omega$ for $p$
polarization at the incidence angle $\protect\vartheta=75^{\circ}$. Local and nonlocal 
cases are
represented by the dashed and solid lines, respectively.
Parameters were chosen as in Fig. \ref{fig2}.} \label{fig3}
\end{figure}

\noindent To verify the procedure we recalculated the absorptance
with $ \gamma =10^{-3}$ and the Fermi velocity $v_{F}=0.85\cdot
10^{8}\;cm/s$ (potassium) to compare with the same calculations in
Ref. \cite{KF1}. The results are presented in Figs. \ref{fig2} and
\ref{fig3}. Absorptance at the normal incidence $ \vartheta =0$
($Q=0$) is shown in Fig \ref{fig2}. In this case both
polarizations give the same result. The nonlocal case is presented
by the solid line. The absorptance in the local limit calculated
with the impedances (\ref {loc}) is given by the dashed line. The
usual increase in the absorptance can be seen at low frequencies
$\Omega \sim 10^{-3}$ due to the anomalous skin effect. For the
incidence angle $\vartheta =75^{\circ }$ the absorptance of the
$p$ polarized wave is shown in Fig. \ref{fig3}. In this case there
is an additional peak in absorptance at higher frequencies $\Omega
\sim 0.1$. It appears only for $p$ polarization; the $s$
polarization behaves similar to the case $\vartheta =0$. At
smaller $\gamma $ both of the peaks become much more significant.
These results are in full agreement with those of Kliewer and
Fuchs \cite{KF3}.

\subsection{Evanescent fields}

The fluctuating currents in the plates are the sources of
fluctuating electromagnetic fields responsible for the Casimir
force.  The typical separation between bodies in the Casimir force
experiments is smaller than the wavelength $\lambda $ of visible
light. If we consider one plate as an emitter and the other one as
a receiver, then for significant part of the spectrum contributing
in the force the receiver will be in the near field zone of the
emitter. In this case the propagating field radiated by the
emitter will be small in comparison with the evanescent field
which exists around the emitter at the distances $\sim \lambda $.
The well known example of such an emitter is the Hertz dipole. At
small distances from the dipole $ \omega r/c\ll 1$ one can neglect
the retardation and the field around the source is just the field
of the static dipole decaying as $1/r^{2} $. When the force is
calculated using the Green function method \cite{LP9}, the Green
function is exactly the dipole field modified by the presence of
the plates. The planar geometry of the problem makes it preferable
to expand the dipole field on the plane waves. The plane waves
obeying the condition $ \omega ^{2}/c^{2}<q^{2}$ do not propagate
in the gap because the normal component of the wave vector is pure
imaginary.

There were some speculations in the literature (see, for example,
\cite{GKM} ), inspired by the problem with the temperature
correction to the Casimir force, that for the evanescent fields,
the standard expressions for the Fresnel reflection coefficients
should be modified. In this connection we have to stress that the
evanescent fields are the subject of the near-field optics
\cite{Cour} (see also \cite{WJ} for a review), where standard
electrodynamic approaches are used. Additionally, the longitudinal
dielectric function can be probed in the evanescent range by the
scattering of a beam of charged particles or fast electrons from
the material \cite{Book3}. In this way the function $ Im
\{1/\varepsilon _{l}({\bf k},\omega )\}$ can be directly extracted
from the experiment, where ${\bf k}$ is connected with the
momentum and $\omega $ with the energy losses of the charged
particles. No necessity for modification of the standard
electrodynamics was noted so far. A consistent way for the
description of evanescent fields is just the analytic continuation
of the Eqs. (\ref{Zs}), (\ref{Zp}), (\ref {etdef})-(\ref{udef}),
and Eq. (\ref{refl}) to the range $\omega ^{2}/c^{2}<q^{2} $.

Originally the Lifshitz formula for the Casimir force was written
as an integral over real frequencies $\omega $ \cite{Lif}. In this
representation one has to calculate first the integral over the
variable $p=\sqrt{ 1-(cq/\omega )^{2}}$ in the range $0<p<1$
(propagating fields) and then integrate over the imaginary axis
$p=i\left| p\right| $ from zero to infinity (evanescent fields).
So the propagating and evanescent fields were clearly separated.
The alternative representation of the same formula \cite {LP9} is
more popular because of faster convergence of the integrals. In
this case the integration is done over the imaginary frequencies
$\omega =i\zeta $ but the inner integral over $p=\sqrt{1+(cq/\zeta
)^{2}}$ is calculated from $1$ to $\infty $. Formally we are
always in the evanescent domain because at imaginary frequencies
the normal component of the wave vector is pure imaginary
$k_{z}=i\sqrt{\zeta ^{2}/c^{2}+q^{2}}$. \ For this reason we will
not investigate specially the domain $q^{2}>\omega ^{2}/c^{2}$
making the analytic continuation on $q$ but instead we will make
the analytic continuation to imaginary frequencies.  This
procedure is well defined for the response functions which are
analytical in the upper half of complex plane $\omega $. In the
electrodynamics the response functions are the components of the
Green function $\varepsilon _{l}^{-1}\left( {\bf k} ,\omega
\right) $ and $\left[ \left( \omega ^{2}/c^{2}\right) \varepsilon
_{t}\left( {\bf k},\omega \right) -k^{2}\right] ^{-1}$ but not the
dielectric functions themselves \cite{Mar,Kir}. Exactly these
expressions take part in the impedances (\ref{Zs}) and (\ref{Zp})
and, therefore, the impedances can be considered as analytical
functions of $\omega $.

Using the dimensional variables (\ref{var}) the impedances at
imaginary frequencies ($\Omega \rightarrow i\Omega $) can be
written as

\begin{equation}
Z_{s}\left( Q,\Omega \right) =\frac{2}{\pi }\frac{\Omega }{Q}%
\int\limits_{0}^{\infty }\frac{\cosh \chi \,d\chi }{\cosh ^{2}\chi +\frac{%
\Omega ^{2}}{Q^{2}}\varepsilon _{t}\left( \Omega ,v\right) },
\label{Zsim}
\end{equation}

\begin{equation}
Z_{p}\left( Q,\Omega \right) =\frac{2}{\pi }\frac{\Omega }{Q}%
\int\limits_{0}^{\infty }\frac{d\chi }{\cosh \chi }\left[ \frac{1}{\frac{%
\Omega ^{2}}{Q^{2}}\varepsilon _{l}\left( \Omega ,v\right)
}+\frac{\cosh ^{2}\chi \,-1}{\cosh ^{2}\chi +\frac{\Omega
^{2}}{Q^{2}}\varepsilon _{t}\left( \Omega ,v\right) }\right] .
\label{Zpim}
\end{equation}

\noindent Here we introduced a new variable of integration $\chi $
which is defined by the relation $k_{z}=k_{x}\sinh \chi $. For the
dielectric functions at imaginary frequencies one finds

\begin{equation}
\varepsilon _{l}\left( \Omega ,v\right) =1+\chi _{IB}\left( \Omega \right) +%
\frac{f_{l}\left( v\right) }{\Omega \left( \Omega +\gamma \right)
},\quad
f_{l}\left( v\right) =\frac{3}{v^{2}}\frac{v-\arctan v}{v+\frac{\gamma }{%
\Omega }\left( v-\arctan v\right) },  \label{Elim}
\end{equation}

\begin{equation}
\varepsilon _{t}\left( \Omega ,v\right) =1+\chi _{IB}\left( \Omega \right) +%
\frac{f_{t}\left( v\right) }{\Omega \left( \Omega +\gamma \right)
},\quad f_{t}\left( v\right) =\frac{3}{2v^{3}}\left[ -v+\left(
1+v^{2}\right) \arctan v\right] ,  \label{Etim}
\end{equation}

\begin{equation}
v=\frac{v_{F}}{c}\frac{Q}{\Omega +\gamma }\cosh \chi .
\label{vdefim}
\end{equation}

\noindent These formulas are used for numerical calculations of
the impedances. They have to be compared with the classical
expressions in the local limit which follows from Eq. (\ref{loc})
after the change to imaginary frequencies:

\begin{equation}
Z_{s}^{loc}=\frac{1}{\sqrt{\varepsilon \left( \Omega \right) +\frac{Q^{2}}{%
\Omega ^{2}}}},\quad Z_{p}^{loc}=\frac{\sqrt{\varepsilon \left(
\Omega
\right) +\frac{Q^{2}}{\Omega ^{2}}}}{\varepsilon \left( \Omega \right) }%
,\quad \varepsilon \left( \Omega \right) =1+\chi _{IB}\left( \Omega \right) +%
\frac{1}{\Omega \left( \Omega +\gamma \right) }.  \label{localim}
\end{equation}

\begin{figure}
\includegraphics[width=8.6cm]{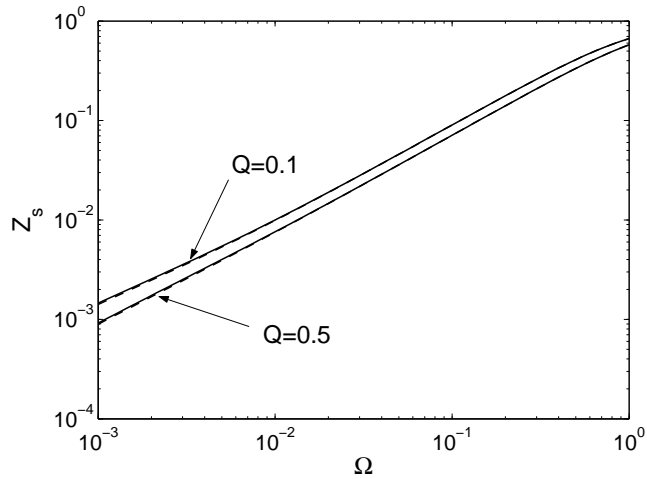}
\caption{Numerically calculated impedance $Z_s$ as a function of
dimensionless frequency $\Omega$ for two values of the
dimensionless wave numbers along the plate $Q$. Solid line
describes nonlocal calculations; the dashed line present the local
case. Maximal deviation between local and nonlocal curves is about
2\%. Gold parameters were used for calculation:
$\protect\gamma=3\cdot10^{-3}$, $v_F=1.4\cdot10^8\ cm/s$.}
\label{fig4}
\end{figure}

The numerical result for $Z_{s}$ is shown in Fig. \ref{fig4} as a
function of $\Omega $ for two values of $Q$. All calculations were
performed for the parameters corresponding to gold at room
temperature: $\gamma =3\cdot 10^{-3} $, $v_{F}=1.4\cdot
10^{8}\;cm/s$, $\omega _{p}=1.37\cdot 10^{16}\;rad/s$. The
impedance of the local theory is presented by the dashed lines.
One can see that the nonlocal effect is very small for this
polarization. The largest deviation from the local curves is just
about of 2\%. Obviously the $s$ polarization cannot produce
significant nonlocal correction to the Casimir force.

\begin{figure}
\includegraphics[width=8.6cm]{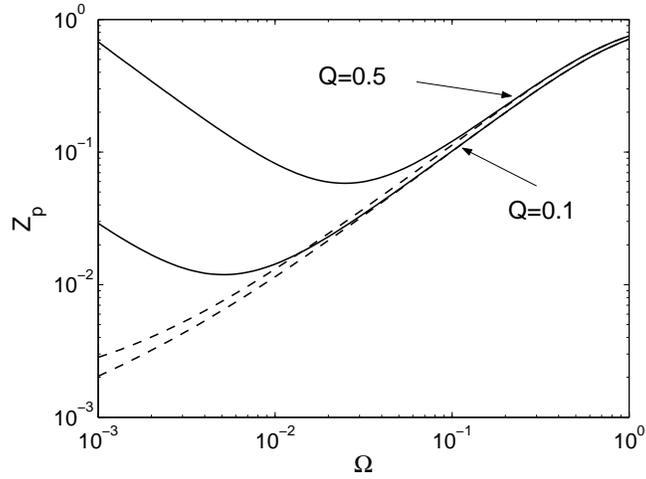}
\caption{Numerically calculated impedance $Z_p$ as a function of
frequency $ \Omega$ for two values of the wave numbers along the
plate $Q$. Nonlocal and local cases are shown by the solid and
dashed lines, respectively.} \label{fig5}
\end{figure}

A different situation is realized for $p$ polarization,  as
shown in Fig. \ref{fig5}. One can see that there is a significant
difference between the local and nonlocal cases. The deviation
increases with frequency decrease and becomes larger for larger
$Q$. This behavior has deep physical meaning, as explained below,
and can appear only for the evanescent fields. Since in both cases
the deviations from the local case are in the low frequency range,
we analyze this limit analytically.

\subsection{Low frequency behavior of impedances}

At low frequencies $\Omega \lesssim \gamma $, the variable $v$
defined by Eq. (\ref {vdefim}) can be large if $\gamma \lesssim
v_{F}/c\approx 4.7\cdot 10^{-3}$. Let us consider the impedances
in the limit $v\gg 1$. In this limit the functions $f_{l}\left(
v\right) $ and $f_{t}\left( v\right) $ in Eq. (\ref{Elim} ) and
Eq. (\ref{Etim}) behave as

\begin{equation}
f_{l}\left( v\right) \approx \frac{3}{v^{2}}\frac{\Omega }{\Omega +\gamma }%
,\quad f_{t}\left( v\right) \approx \frac{3\pi }{4v},\quad v\gg 1.
\label{flim}
\end{equation}

\noindent In the transverse dielectric function $\varepsilon _{t}$
one can neglect $1+\chi _{IB}\left( \Omega \right) $ since the
third term behaves as $1/\Omega $ at low frequencies. It gives for
$\varepsilon _{t}\left( \Omega ,v\right) $

\begin{equation}
\varepsilon _{t}\left( \Omega ,v\right) \approx \frac{4\pi }{3}\frac{c}{v_{F}%
}\frac{1}{Q\cosh \chi }\frac{1}{\Omega }.  \label{etappr}
\end{equation}

\noindent For the the longitudinal function $\varepsilon _{l}$ the
phenomenological susceptibility $\chi _{IB}\left( \Omega \right) $
again is negligible because it is responsible for the interband
transitions at much higher frequencies but we cannot neglect the
unit since the third term in (\ref{Elim}) does not depend on
frequency at all and not necessarily large.  For
$\varepsilon _{l}\left( \Omega ,v\right) $ one find

\begin{equation}
\varepsilon _{l}\left( \Omega ,v\right) \approx 1+3\left( \frac{c}{v_{F}}%
\frac{1}{Q\cosh \chi }\right) ^{2}=1+3\left( \frac{\omega _{p}}{v_{F}}\frac{1%
}{k}\right) ^{2}.  \label{elappr}
\end{equation}

\noindent This expression describes the Thomas-Fermi screening of
the longitudinal electric field. It has to be true \cite{KF2} at
$\Omega <\gamma $ and $k$ much smaller than the Fermi wave number
$k_{F}$ that is the applicability range of the Thomas-Fermi
approximation. The latter condition $ k\ll k_{F}$ is also the
condition for applicability of the Boltzmann equation.

Substituting (\ref{etappr}) and (\ref{elappr}) in Eqs.
(\ref{Zsim}) and (\ref {Zpim}) one finds for the impedances

\begin{equation}
Z_{s}=\frac{\Omega }{Q}F(b),  \label{Zsanom}
\end{equation}

\begin{equation}
Z_{p}=\frac{Q}{\Omega }\frac{1}{\sqrt{1+3\left( c/v_{F}Q\right) ^{2}}}+\frac{%
\Omega }{Q}G(b)\approx \frac{1}{\sqrt{3}}\frac{Q^{2}}{\Omega }\frac{v_{F}}{c}%
+\frac{\Omega }{Q}G(b),  \label{Zpanom}
\end{equation}

\noindent where the functions $F\left( b\right) $ and $G\left(
b\right) $ are defined as

\begin{equation}
F(b)=\frac{2}{\pi }\int\limits_{0}^{\infty }d\chi \frac{\cosh ^{2}\chi }{%
\cosh ^{3}\chi +b^{3}},\quad G\left( b\right) =\frac{2}{\pi }%
\int\limits_{0}^{\infty }d\chi \frac{\sinh ^{2}\chi }{\cosh ^{3}\chi +b^{3}%
},  \label{FGdef}
\end{equation}

\begin{equation}
b=\frac{1}{Q}\left( \frac{3\pi }{4}\frac{c}{v_{F}}\Omega \right)
^{1/3}. \label{bpar}
\end{equation}

\begin{figure}
\includegraphics[width=8.6cm]{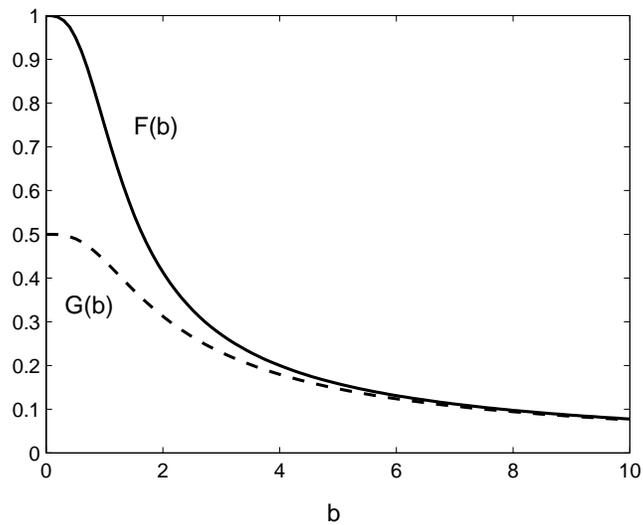}
\caption{Numerically calculated functions $F(b)$ and $G(b)$.}
\label{fig6}
\end{figure}

\noindent The functions $F(b)$ and $G(b)$ can be found explicitly
but the result is cumbersome and inconvenient for analysis. For
this reason we calculated the integrals in Eq. (\ref{FGdef})
numerically presenting explicitly only the asymptotics at $b\ll 1$
and $b\gg 1$. The functions $F(b)$ and $G(b)$ are shown in Fig.
\ref{fig6}. The asymptotic behavior of these functions is

\[
F(b)=\left\{
\begin{array}{c}
1-\frac{4}{3\pi }b^{3},\quad b\ll 1 \\
\frac{4}{3\sqrt{3}}\frac{1}{b}+\frac{1}{\pi b^{3}}\left( \ln
2b-1/2\right) ,\quad b\gg 1
\end{array}
\right.
\]

\begin{equation}
G(b)=\left\{
\begin{array}{c}
\frac{1}{2}-\frac{4}{15\pi }b^{3},\quad b\ll 1 \\
\frac{4}{3\sqrt{3}}\frac{1}{b}-\frac{1}{\pi b^{3}}\left( \ln
2b+1/2\right) ,\quad b\gg 1
\end{array}
\right.  \label{asympt}
\end{equation}

The known result for the Leontovich impedance for the strong
anomalous skin effect \cite{RS,LP10,Abr} is easily reproduced if
we take in the equations above the limit $Q\rightarrow 0$. In this
limit the parameter $b$ goes to infinity and the contribution
of the transverse dielectric function is the same for both
polarization: $F(b)=G(b)=4/3\sqrt{3}b$. The contribution from
$\varepsilon _{l}$ in $Z_{p}$ disappears in the limit $Q\rightarrow 0$.
Hence, the impedances will coincide with each other and they are
given by the classical expression for the strong anomalous skin
effect continued to imaginary frequencies

\begin{equation}
Z_{s}(0,\Omega )=Z_{p}\left( 0,\Omega \right) =Z\left( \Omega \right) =\frac{%
4}{3\sqrt{3}}\left( \frac{4}{3\pi }\frac{v_{F}}{c}\Omega
^{2}\right) ^{1/3}. \label{Leont}
\end{equation}

\noindent However, if $Q$ is nonzero  there is a small
enough frequency where $b$ is not large anymore and Eq.
(\ref{Leont}) is not applicable. When $\Omega $ is so small that $b\ll
1$ the impedance $Z_{s}$ approaches the limit $\Omega /Q$. The
same limit is realized for the local impedance $Z_{s}^{loc}$ in
Eq. (\ref{localim}) at very low frequencies when one
can neglect $\varepsilon \left( \Omega \right) $ in comparison with $%
Q^{2}/\Omega ^{2}$.

For $p$ polarization at nonzero $Q$ the contribution of
$\varepsilon _{t} $ in the impedance decreases with $\Omega $ but
the contribution of $ \varepsilon _{l}$ increases as $1/\Omega $
(see Eq. (\ref{Zpanom})) and dominates in $Z_{p}$ at low
frequencies. It is in agreement with our numerical calculations in
Fig. \ref{fig5}. Indeed this is the result of the Thomas-Fermi
screening. The same effect is not realized for the propagating
fields. In this case the ratio $Q/\Omega =\sin \vartheta \leq 1$
is restricted. Since $v_{F}/c=4.7\cdot 10^{-3}$ is small, the
variable $u$ (\ref {udef}) is small nearly everywhere in the
integration range and the function $f_{l}\left( u\right) \approx
1$. Therefore the local limit is realized instead for the
longitudinal contribution in $Z_{p}$.

The behavior of the impedance for the $s$ polarization at low
frequencies is a sensitive matter for the temperature correction
to the Casimir force. One of us (VBS) in collaboration with M.
Lokhanin analyzed this problem \cite{SL4} with the Leontovich
impedance (\ref{Leont}). As follows from the discussion above this
analysis has to be reconsidered taking into account not only
different behavior of $Z_{s}$ at very low frequencies but also
significant deviation of $Z_{p}$ from the local impedance in this
range.

\section{Nonlocal correction to the Casimir force\label{sec3}}

In this section we are going to estimate the correction to the
Casimir force due to nonlocal effects at frequencies smaller than
$\omega _{p}$. The restriction on frequency is connected with the
use of the Boltzmann approximation for the dielectric functions
(\ref{Elim}), (\ref{Etim}). In this approximation we cannot
describe the plasmon excitations. Of course, one could use more
general dielectric functions like those in the
self-consistent-field approximation \cite{KF2} to analyze all the
nonlocal effects. However, we think it is reasonable to separate
the effects of different physical origin. Influence of the plasmon
excitations on the Casimir force has been already evaluated
\cite{EVM} using the hydrodynamic approximation for the
longitudinal dielectric function, but the correction to the force
due to the anomalous skin effect never has been calculated before.
Only specific questions concerning the temperature correction have
been addressed in the literature \cite{SL4}. By anomalous skin
effect we refer not only to the strong anomalous skin effect that
is realized when the electron mean free path is larger than the
field penetration depth, but to all the nonlocal effects that
happen at frequencies smaller than $\omega _{p}$.

We will consider only the force in the zero temperature limit. Thus,
the Casimir force will be calculated without the
temperature correction but all the other parameters characterizing
the material, especially the relaxation frequency $\omega _{\tau
}$, will be kept at finite temperature. The results of the
previous section for the impedances $Z_{s}$ and $Z_{p}$ are
important for the temperature correction problem but this question
will be considered elsewhere.

It is known that when a metal is described by the surface
impedances, the Lifshitz formula for the Casimir force \cite{LP9}
remains essentially the same \cite{MT,EVM} as when the metal is
described by the local dielectric function. Only the reflection
coefficients have to be expressed via the impedances. At nonzero
temperature the Lifshitz formula includes summation over the
Matsubara frequencies $\zeta _{n}$, defined for our dimensionless
frequency as

\begin{equation}
\Omega _{n}=\frac{\zeta _{n}}{\omega _{p}}=\frac{2\pi nkT}{\hbar \omega _{p}}%
.  \label{Mats}
\end{equation}

\noindent To get the Casimir force at $T=0$ we have to integrate
over the continuous variable $\Omega $. In the dimensionless
variables $\Omega $ and $ Q$ the Casimir force between two
metallic plates separated by the distance $ a $ at $T=0$ is

\[
F_{pp}\left( a\right) =-\frac{\hbar c}{2\pi ^{2}\delta ^{4}}%
\int\limits_{0}^{\infty }d\Omega \int\limits_{0}^{\infty }dQQ\sqrt{\Omega
^{2}+Q^{2}}\left[ \left( r_{s}^{-2}\exp \left( 2d\sqrt{\Omega ^{2}+Q^{2}}%
\right) -1\right) ^{-1}+\right.
\]

\begin{equation}
\left. \left( r_{p}^{-2}\exp \left( 2d\sqrt{\Omega ^{2}+Q^{2}}\right)
-1\right) ^{-1}\right] ,  \label{forcepp}
\end{equation}

\noindent where

\begin{equation}
d=\frac{a}{\delta },\quad \delta =\frac{c}{\omega _{p}}\approx 21.88\;nm.
\label{depth}
\end{equation}

\noindent Here $\delta $ is the penetration depth for gold. The
reflection coefficients follows from Eq. (\ref{refl}) after
continuation to imaginary frequencies

\begin{equation}
r_{s}=\frac{\Omega -\sqrt{\Omega ^{2}+Q^{2}}Z_{s}\left( \Omega ,Q\right) }{%
\Omega +\sqrt{\Omega ^{2}+Q^{2}}Z_{s}\left( \Omega ,Q\right) },\quad r_{p}=%
\frac{\sqrt{\Omega ^{2}+Q^{2}}-\Omega Z_{p}\left( \Omega ,Q\right) }{\sqrt{%
\Omega ^{2}+Q^{2}}+\Omega Z_{p}\left( \Omega ,Q\right) }.  \label{reflim}
\end{equation}

\noindent The impedances $Z_{s,p}$ are calculated according to Eqs. (\ref{Zsim}%
)-(\ref{vdefim}).

The force between a sphere and a plane can be calculated with the help of the
proximity force approximation \cite{DA} which gives the following expression

\[
F_{sp}\left( a\right) =\frac{\hbar cR}{2\pi \delta ^{3}}\int\limits_{0}^{%
\infty }d\Omega \int\limits_{0}^{\infty }dQQ\left[ \ln \left(
1-r_{s}^{2}\exp \left( -2d\sqrt{\Omega ^{2}+Q^{2}}\right) \right) +\right.
\]

\begin{equation}
\left. \ln \left( 1-r_{p}^{2}\exp \left( -2d\sqrt{\Omega ^{2}+Q^{2}}\right)
\right) \right] ,  \label{forcesp}
\end{equation}

\noindent where $R$ is the radius of the sphere.

\subsection{Numerical procedure}

\begin{figure}
\includegraphics[width=8.6cm]{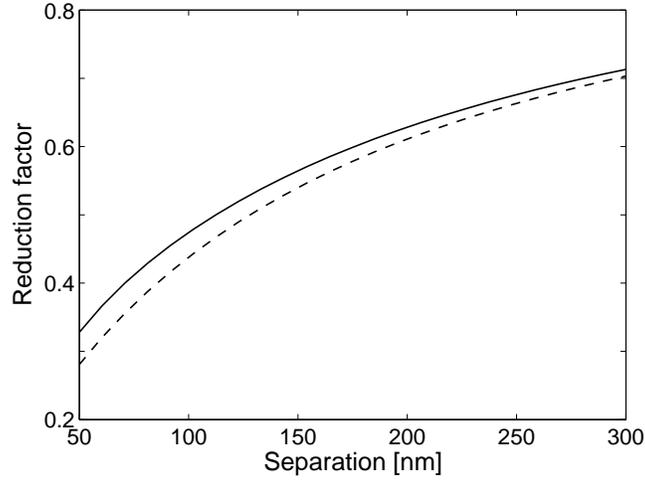}
\caption{The reduction factor $F_{pp}(a)/F_{c}(a)$ in the local
case as a function of the separation $a$ calculated with the
handbook data (solid line) and with the Drude model for the
dielectric function (dashed line).} \label{fig7}
\end{figure}

First we calculate the force between parallel plates
$F_{pp}^{Drude}(a)$ in the local limit with the Drude dielectric
function and local impedances (\ref{localim}). The actual
calculations were performed for the dimensionless relaxation
frequency (see (\ref{var})) $\gamma =4\cdot 10^{-3}$. This value
is the best fit \cite{LR} of the handbook optical data for gold
\cite{HB1} at low frequencies.  In Fig. \ref{fig7} we show the
reduction factor $\eta(a)$  defined as the ratio of the calculated
force to the original Casimir force (\ref{Fc}); this is
\begin{equation}
\eta (a)=\frac{F_{pp}(a)}{F_{c}(a)}.  \label{reduc}
\end{equation}

The force calculated using the Drude model  (dashed line) is
smaller than that calculated using the handbook optical data for
gold (solid line). The solid line coincides with the reduction
factor given in Ref. \cite{LR}.

The nonlocal correction is calculated without the empirical
susceptibility  $\chi _{IB}\left( \omega \right) $ introduced in
Eqs. (\ref{etdef}) and (\ref{eldef}) so we have to remember that
the relative nonlocal correction will be smaller than the
calculated one on the value of the order of the relative
difference between the curves in Fig. \ref{fig7} (8\% at $a=100$\
nm).

\begin{figure}
\includegraphics[width=8.6cm]{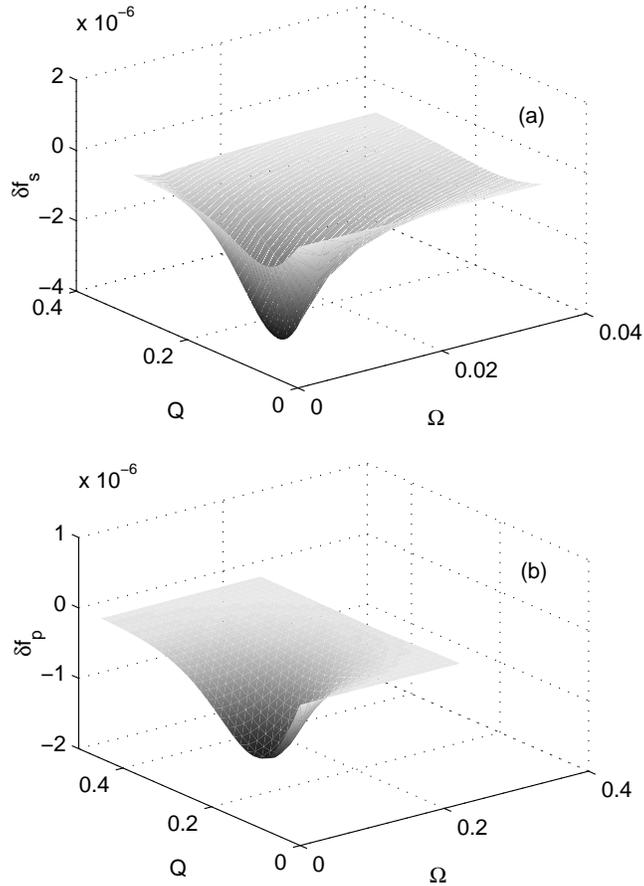}
\caption{Integrands for $s$ polarization (a) and for $p$
polarization (b). Note different scales in $\Omega$ axes.}
\label{fig8}
\end{figure}

The force calculation in the nonlocal case is quite complicated
because one has to make three integrations with high precision,
one to calculate the impedances and two to calculate the force. It
is much more easy to calculate not the force itself but integrate
the difference between nonlocal and local integrands. In this case
there is no need to perform high precision calculation of the
integrals since we have to know the correction due to nonlocality
with the precision of about of 10\%. Actual calculation of the
difference

\begin{equation}
\delta F_{pp}\left( a\right) =F_{pp}^{nonloc}\left( a\right) -F_{pp}^{loc}(a)
\label{cordef}
\end{equation}

\noindent were made with the relative accuracy of 1\%, while the
impedances (\ref{Zsim}), (\ref{Zpim}) were calculated with the
relative precision of $10^{-6}$. The integrands for $s$ and $p$
polarizations defined as

\begin{equation}
\delta f_{s,p}=Q\sqrt{\Omega ^{2}+Q^{2}}\left[ \left( r_{s,p}^{-2}\exp
\left( 2d\sqrt{\Omega ^{2}+Q^{2}}\right) -1\right) ^{-1}-(r_{s,p}\rightarrow
r_{s,p}^{loc})\right]  \label{integrand}
\end{equation}

\noindent are presented for $a=275\;nm$ in Fig. \ref{fig8}(a) and
Fig. \ref {fig8}(b), respectively. Both of them are negative as it
should be, since the force decreases due to the nonlocal effects.
It is interesting to notice that $\delta f_{s}$ is nonzero in a
very narrow range of small $\Omega $.   In contrast, the integrand
for  $p$ polarization  $ \delta f_{p}$  is nonzero in a broader
range of $\Omega $ (pay attention on different scales in figures
(a) and (b)). Nonlocal effects are always significant in a wider
range of $Q\lesssim 1$. With the decrease of separation $a$ the
integrand for $p$ polarization increases in the magnitude and
becomes wider in both directions $\Omega $ and $Q$. The integrand
for $s$ polarization decreases in  magnitude and widens only in
$Q$ direction. Thus, the contribution of $s$ polarization in the
force correction is always smaller than that for the $p$
polarization.

\begin{figure}
\includegraphics[width=8.6cm]{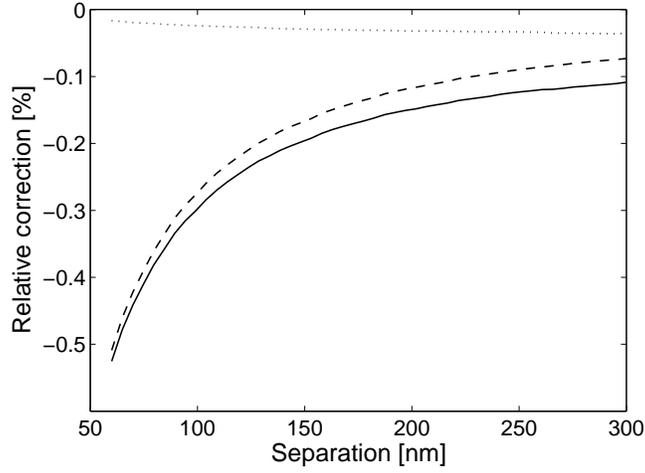}
\caption{The relative correction to the force due to nonlocal
effects for the plate-plate geometry. The solid line presents the
resulting correction. The dashed line gives the contribution of
the $p$ polarization and the dotted line gives the contribution of
the $s$ polarization. Gold parameters were used for calculations.}
\label{fig9}
\end{figure}

The results for the relative correction $\delta F_{pp}\left(
a\right) /F_{pp}^{Drude}\left( a\right) $ due to the nonlocal
effects are presented in Fig. \ref{fig9}. The solid line gives the
resulting correction, while the dashed and the dotted lines
represent the contributions of $p$ and $s$ polarizations,
respectively. One can see that the correction is small but not
negligible. Contribution of $s$ polarization increases when
$\gamma $ becomes smaller, but even for $\gamma =10^{-5}$ this
contribution is still on the level of 0.2\%. We can see that large
deviation of the impedance for $ p$ polarization from the local
one that happens at low frequencies is not very significant for
the force. This is because in the reflection coefficient $r_{p}$
the impedance enter as $\Omega Z_{p}$ so that $1/\Omega $ behavior
of $Z_{p}$ is suppressed in the reflection coefficient.

\begin{figure}
\includegraphics[width=8.6cm]{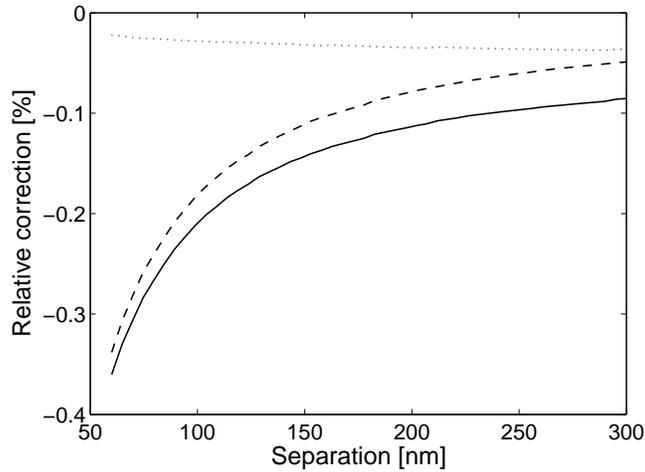}
\caption{Same as Fig. \ref{fig9} but for the sphere-plate
geometry.} \label{fig10}
\end{figure}

Similar calculations were made for the sphere-plate geometry. The
relative correction $\delta F_{sp}\left( a\right)
/F_{sp}^{Drude}\left( a\right) $ together with the separate
contributions of $p$ and $s$ polarizations is shown in Fig.
\ref{fig10}. The behavior of the curves is quite similar to that
for the plate-plate geometry. Only the absolute magnitude of the
correction is  smaller.

\section{Discussion\label{sec4}}

The theory described in Sec. \ref{sec2} provides a solid ground
for the impedance approach in the Casimir force calculation.
Specifically it allows correctly to take into account nonlocal
connection between the displacement and electric fields. In this
paper we restricted ourselves by the nonlocal effects happening at
frequencies smaller than $\omega _{p}$. This restriction is due to
used Boltzmann approximation for the nonlocal dielectric functions
(\ref{etdef})-(\ref{udef}). However the equations for the
impedances (\ref{Zs}), (\ref{Zp}) are much more general. For
specular electron reflection off the surface these equations are
true for arbitrary dielectric functions $\varepsilon _{t}\left(
{\bf k},\omega \right) $ and $ \varepsilon _{l}\left( {\bf
k},\omega \right) $ with the only condition that these functions
exist. Therefore, all nonlocal effects can be described on the
same basis. In particular, for metals the most general dielectric
functions for a free-electron gas were found in the
self-consistent-field (or Lindhard) approximation with the
necessary modifications to include a finite relaxation time
\cite{KF2}. In this approximation $\varepsilon _{t}\left( k,\omega
\right) $ has the following form
\[
\varepsilon _{t}\left( k,\omega \right) =1-\frac{\omega
_{p}^{2}}{\omega \left( \omega +i\omega _{\tau }\right)
}f_{t}\left( u,z\right) ,
\]

\noindent where

\[
f_{t}(u,z)=\frac{3}{8}\left( z^{2}+3u^{-2}+1\right) -\frac{3}{32z}\left\{ %
\left[ 1-\left( z-u^{-1}\right) ^{2}\right] ^{2}\ln \left( \frac{z-u^{-1}+1}{%
z-u^{-1}-1}\right) +\right.
\]

\begin{equation}
\left. \left[ 1-\left( z+u^{-1}\right) ^{2}\right] ^{2}\ln \left( \frac{%
z+u^{-1}+1}{z+u^{-1}-1}\right) \right\} .  \label{etscf}
\end{equation}

\noindent Here $u$ defined as before by Eq. (\ref{udef}) and $z$ is $%
z=k/2k_{F}$. The longitudinal dielectric function has a little bit
more complicated form:

\[
\varepsilon _{l}\left( k,\omega \right) =1+(\varepsilon _{w}-1)\left[ 1+i%
\frac{\omega _{\tau }}{\omega }-\frac{i}{2u}\frac{\omega _{\tau }}{\omega }%
\ln \left( \frac{1+u}{1-u}\right) \right] ^{-1},
\]

\[
\varepsilon _{w}=1+\frac{3\omega
_{p}^{2}}{k^{2}v_{F}^{2}}f_{l}\left( u,z\right) ,
\]

\[
f_{l}\left( u,z\right) =\frac{1}{2}+\frac{1}{8z}\left\{ \left[
1-\left( z-u^{-1}\right) ^{2}\right] \ln \left(
\frac{z-u^{-1}+1}{z-u^{-1}-1}\right) +\right.
\]

\begin{equation}
\left. \left[ 1-\left( z+u^{-1}\right) ^{2}\right] \ln \left( \frac{%
z+u^{-1}+1}{z+u^{-1}-1}\right) \right\} . \label{elscf}
\end{equation}

\noindent All the other approximations for the free-electron gas
can be found from (\ref{etscf}), (\ref{elscf}) in definite limit
cases. For example, the Boltzmann approximation
(\ref{etdef})-(\ref{udef}) follows from (\ref{etscf}),
(\ref{elscf}) in the limit $z\rightarrow 0$. These expressions for
the dielectric functions allow to perform detailed investigation
of the high frequency region $\omega \gtrsim \omega _{p}$ which
gives more significant contribution in the Casimir force due to
excitation of the propagating charge density waves in the metal
\cite{EVM}.

We considered here only specular electron reflection off the metal
surface. It is justified for the AFM experiment \cite{Moh2} where
the root mean square (rms) roughness of the surface ($1\;nm$) was
much smaller than the mean free path ($30\;nm$). However, in the
MEMS experiments \cite {Chan,Decca1,Decca2} the rms roughness was
comparable with the mean free path and approximation of specular
reflection fails. In this case the diffuse reflection of electrons
off the surface is more suitable. For the diffuse reflection the
impedances are not represented by the Eqs. (\ref{Zs} ), (\ref{Zp})
anymore. Instead one has to use the impedances for the diffuse
reflection \cite{KF3}. There is no problem with $Z_{s}$ which is
expressed via $\varepsilon _{t}\left( k,\omega \right) $ but
situation with $Z_{p}$ is much more complicated. This occurs
because of the destruction of translational invariance in the
direction normal to the surface \cite{KF3}. Although it is
possible to calculate both of the impedances in the diffuse case,
we do not think it is reasonable to do for the anomalous skin
effect. This is because the nonlocal correction is smaller than
the uncertainty in the Casimir force due to the roughness. The
roughness correction to the force is usually evaluated using the
proximity force approximation (see, for example, \cite{Decca2}).
Recently it has been pointed out \cite{GLNR} that this approach is
valid only for long wavelength deformations of the plates. The
real surfaces of deposited gold films always roughed on very
different scales \cite{OASA} and the short wavelengths will bring
uncertainty in the estimate of the force.

The impedances (\ref{Zs}), (\ref{Zp}) with the nonlocal dielectric
functions (\ref{etdef})-(\ref{udef}) are well known in the optics
of metals but in this paper we considered them in the near field
range where the nonlocal effects were unexplored. In this sense
the Casimir force is a unique problem. Significant contribution in
the force comes from fluctuating fields in the near field region.
Since the force has to be predicted with high precision, it is
important to take into account the nonlocal effects. Though we
have found here that the anomalous skin effect gave observable but
minor correction to the force, the other nonlocal effects, such as
plasmon excitation, can give more significant correction. This
paper just provide a regular basis for the calculations of this
kind.

\section{Conclusions\label{sec5}}

A complete calculation of the Casimir force that can be accurately
compared with experiments requires a full optical characterization
of the involved materials. This is complicated due to the various
factors that can modify the optical properties. In this work we
described a systematic way to take into account the nonlocal
effects in the material. It was stressed that, in general, a metal
had to be described with two different surface impedances
corresponding to $s$ and $p$ polarizations and these impedances
depend not only on frequency but also on the wave vector along the
metal surface. As a specific problem we considered the correction
to the Casimir force in the region of the anomalous skin effect
($\omega<\omega_p$). This region is characterized by the nonlocal
dielectric functions (longitudinal and transverse) that can be
obtained in the Boltzman approximation. The impedances are
completely defined by these functions.

It was demonstrated that the exact impedances were differed from
the approximate Leontovich impedance. The latter one caused
confusion in the literature so our analysis resolved the problem
and gave a proper description of the impedance approach in the
Casimir force calculation. It was emphasized that the significant
contribution in the force came from the evanescent fields. For
these fields the impedances can be found by the analytic
continuation and the procedure is well defined. The contribution
of the nonlocal effects in the impedances was found to be quite
different for propagating and evanescent fields. Specifically for
the evanescent fields the impedance for $p$ polarization deviates
significantly from the local one that is the result of the
Thomas-Fermi screening. For $s$ polarization the nonlocal
contribution in the impedance is more significant for the
propagating fields than for the evanescent ones.

In the impedance approach the Casimir force can be found from the
same Lifshitz formula in which the reflection coefficients are
expressed via the impedances. We calculated the nonlocal
correction to the force in the region of anomalous skin effect at
zero temperature. In spite of significant deviation of $Z_p$ from
the local impedance the nonlocal reflection coefficient $r_p$
deviates from the local one only slightly. For the $s$
polarization the effect is even smaller. For this reason the total
contribution of the nonlocal effects in the Casimir force is on
the level of $0.5\%$ at small separations. It is a minor effect  within the levels
of detectability of present experiments, but smaller than the corrections introduced due
to sample roughness.  

We did not considered in this paper the temperature correction
though it is clear from the analysis of impedances that anomalous
skin effect will be important for the temperature correction. A
new phenomenon observed here is that the reflection coefficient
$r_p$ for $p$ polarization is not going to 1 in the zero frequency
limit $\omega\rightarrow 0$. This behavior is the result of the
Thomas-Fermi screening.

The technic developed in this paper can be applied to calculate
the contribution of the other nonlocal effects such as plasmon
excitation at $\omega\sim\omega_p$. These effects are expected to
give more significant correction to the Casimir force.

\acknowledgements{RE acknowledges the partial support provided by
DGAPA-UNAM projects IN116002, IN117402 and CONACyT project No. 36651-E. VBS is grateful to the Transducer Science and Technology group, Twente University, for
hospitality and acknowledges the support from the Dutch
Technological Foundation.}

\newpage

\end{document}